\documentclass[twocolumn,showpacs,preprintnumbers,amsmath,amssymb,floatfix,pre]{revtex4}
\usepackage{graphicx}
\usepackage{dcolumn}
\usepackage{bm}
\usepackage{color}
\usepackage{psfrag}
\usepackage[german,english]{babel}
\usepackage[latin1]{inputenc}
\usepackage{psfrag}

\newcommand{\zmc}{two-mirror cavity}
\newcommand{\dmc}{three-mirror cavity}
\newcommand{\vmc}{four-mirror cavity}
\newcommand{\tf}{transfer function}

\begin{document}

\preprint{APS/123-QED}
\title{Analysis of a four-mirror cavity enhanced Michelson interferometer}

\author{Andr\'{e} Th\"uring}
\author{Harald L\"uck}
\author{Karsten Danzmann}
\affiliation{Max-Planck-Institut f\"ur Gravitationsphysik (Albert-Einstein-Institut), and Institut f\"ur Gravitationsphysik, Universtit\"at Hannover, Callinstrasse 38, 30167 Hannover, Germany}

\date{\today}

\begin{abstract}
We investigate the shot noise limited sensitivity of a four-mirror cavity enhanced Michelson interferometer. The intention of this interferometer topology is the reduction of thermal lensing and the impact of the interferometers contrast although transmissive optics are used with high circulating powers. The analytical expressions describing the light fields and the frequency response are derived. Although the parameter space has 11 dimensions, a detailed analysis of the resonance feature gives boundary conditions allowing systematic parameter studies.  
\end{abstract}


\maketitle
     
\section{Introduction}
To improve the strain sensitivity of interferometric gravitational wave detectors  advanced interferometer topologies such as Resonant Sideband Extraction (RSE)\cite{miz2} and Signal Recycling\cite{meers} will be realized in the next generation. The sensitivity in the shot noise limited region will be increased by a factor of about 10 over the current detectors by increasing the circulating laser power. Using the Power Recycling technique together with high finesse arm cavities in each interferometer arm and high power lasers, the circulating light power will almost reach the Megawatt regime. But the performance and sensitivity of these topologies strongly depends on the interferometers contrast. Thus, we investigated a topology- the \vmc{} enhanced Michelson- with the potential to minimize the influence of imperfections in the contrast and in addition the effect of thermal lensing although transmissive optics are used with high circulating powers. 

In this paper the investigation of the \vmc{} with respect to the shot noise limited sensitivity is presented. In Section~\ref{analysis} the analytic expressions for the carrier fields and the frequency response of the \vmc{} will be derived. The expressions will be given in analogy to a \zmc{} (Fabry-Perot-Resonator) offering an intuitive understanding of the whole coupled system. Because of the huge parameter range, configurations yielding satisfactory sensitivities are not obvious. Thus, to analyze the \vmc{} systematically some basic assumptions and boundary conditions are necessary. In Section~\ref{ropr} such boundary conditions will be derived from a detailed analysis and visualization of the resonance feature. This analysis reduces the number of free parameters from 11 to 6. Furthermore, it will be shown that the parameters for the second cavity can be chosen to give peak sensitivities at selectable frequencies. These parameters serve as a starting point for all other parameters which drastically reduces  the number of steps in numerical parameter studies. In Section~\ref{study} the dependence on the free parameters is demonstrated for special cases and exemplary shot noise limited sensitivities and properties of the \vmc{} are shown in comparison to Advanced LIGO\cite{ligo}. 

The intention of the future interferometric gravitational wave detectors is the enhancement of the shot noise limited sensitivity by increasing the circulating light powers inside the interferometer. Here, the available laser input field would be ideally exploited if no power is reflected to the interferometers input. This can be realized using impedance matched Fabry-Perot resonators (cavities) in each interferometer arm. Using high finesse resonators would also provide desirable high circulating powers. But the bandwidths of these resonators are very small leading to an unsatisfactory sensitivity in the detection band beyond the arm resonators' bandwidth. To broaden the bandwidth for signal sidebands the RSE scheme\cite{miz2} was proposed. Here, an additional mirror is placed in the interferometer output forming together with the arm cavities' coupling mirrors the extraction cavity. Since this cavity includes the beam splitter, the performance of the sideband extraction strongly depends on the interferometers contrast. If the intra cavity losses (mainly caused by a bad interferometer contrast) become comparable to the transmission of the arm cavities' coupling mirrors, the sideband extraction collapses. To overcome this problem, the power recycling technique\cite{drever-pr} is used. This technique allows the increasing of the coupling mirrors transmission so that the intra cavity losses in the extraction cavity become less significant. To maintain the effective power build up in the arm cavities, the wasted power (now reflected from the overcoupled arm cavities) is recycled in the power recycling cavity (also including the beam splitter) in such a way, that the complete optical configuration is still impedance matched. However, the power recycling leads to relative high optical powers in the recycling cavity. Here the amount of allowable power is limited by the nonzero absorption of the used transmissive optics and their coatings. Due to the effects of thermal lensing\cite{wink,strain} and thermal expansion of the optics surfaces the heating by  optical absorption causes  phase front distortions leading to poor interference quality (critical for the RSE-scheme) at the dark port operating point. This causes higher optical power on the photo detector and therefore higher shot noise. Furthermore, the thermal lensing in the substrates of the arm cavities' coupling mirrors leads to an unstable cavity for the RF-modulation sidebands needed for controlling the interferometers\cite{McClel}. 

The strong dependence on the contrast and the effect of thermal lensing in these advanced interferometers were the motivation for investigating techniques and alternative interferometer topologies to solve these problems. 
The use of alternative substrate materials, for example sapphire, is one solution~\cite{saph}.  Active thermal compensation  provides a further opportunity  to reduce thermally induced phase front distortions\cite{Lueck,law1,law2,blair}. Basically different and  promising is the use of all-reflective interferometer topologies\cite{drever}. At present the application of gratings in future detector topologies and implementation issues are investigated\cite{bunki1,bunki2}. However, the fabrication of appropriate reflection gratings with the dimensions and quality needed for the application in interferometric  gravitational wave detectors still poses a problem.  

\begin{figure}
\includegraphics[scale=0.55]{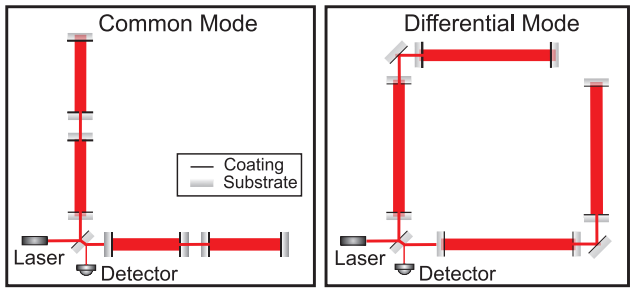}
\caption{\label{ifo}(color online) Two schematics of a \vmc{} enhanced Michelson interferometer: The left shows the unfolded realization. The right setup includes an angle of ninety degrees between the first and third resonators exploiting the quadrupole nature of gravitational waves. Here the sidebands are generated differentially in the first and third resonators.}
\end{figure}

We investigated a topology using additional mirrors in each interferometer arm and not in the input and output. If an additional mirror is placed in each interferometer arm the power build up and the extraction of signal sidebands could be performed without including the beam splitter in a cavity. Thus, the losses at the beam splitter would not limit the performance of the interferometer. But with a \dmc{} the effect of thermal lensing could not be avoided because at least one optical substrate is embedded in a resonator with high circulating power. Also, the resonance condition of carrier and sidebands would not be decoupled. Thus, with a three-mirror arm cavity there would be no way to tune the resonators for carrier and sidebands independently. But if a fourth mirror is placed in the arms, a second long resonator is formed (Figure~\ref{ifo}). This configuration yields resonance states where the circulating light power in the second cavity embedding the optical substrates is  small compared to those of the first and third resonators. In this configuration, the resonance conditions of carrier and sidebands are also coupled. However, the coupling between the first and third resonator can be varied with the effective transmission of the second resonator. Thus, the frequencies of the corresponding resonance doublet of these coupled resonators are tunable. In fact the \vmc{} has a resonance triplet. But the length of the second resonator is chosen to be small (see Section~\ref{ropr}) leading to a high resonance frequency with no effect in the frequency range of interest.    

\section{\label{analysis}Analytical description}
\subsection{Carrier fields}

Since the reason for increasing the circulating light power in advanced topologies is the improvement of the sensitivity in the shot noise limited frequency region, initially the impact of radiation pressure noise is not considered. Moreover, taking radiation pressure noise into account would expand the parameter range as the masses of the mirrors and the input power also influence the frequency response. Thus, it is suitable to select parameter configurations at first with respect to  satisfying shot noise limited sensitivities. After that the selected configurations can be tested for the effect of radiation pressure.

\begin{figure}[h]
\includegraphics[scale=.38]{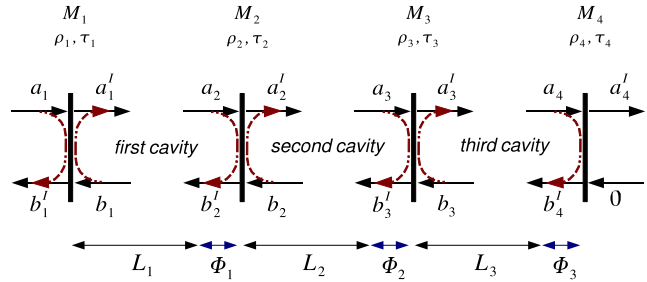}
\caption{\label{notation}(Color online) The notation used in this paper: the figure illustrates the amplitude coupling in a \vmc{}. Here $\rho_j$ and $\tau_j$ are amplitude reflection and transmission of the corresponding mirror M$_j$. The fields coupled at the mirror M$_j$ are denoted $a_j$, $a^\prime_j$ and $b_j$, $b_j^\prime$. The macroscopic lengths of the resonators are labeled with $L_k$. These lengths are assumed to be an integer multiple of half the wavelength corresponding to the carrier light frequency $\omega_0$. Then, the resonance is determined by the tuning (microscopic lengths) $\Phi_k=\omega L_k/c\bmod 2\pi$ (Note that $\omega_0L_k/c$ is 0).}
\end{figure}
If the mirrors are assumed to be ideal (loss free) the carrier fields $C_k$ in the \vmc{} can be  calculated in analogy to a \zmc{}. The enhancement of the input fields $a_k$  inside the corresponding cavities is given by (the notation refers to Figure~\ref{notation})
\begin{subequations}
\begin{eqnarray}
C_3&=&a_3^\prime=\frac{i\tau_3}{1-\rho_3\rho_4\,e^{2i\Phi_3}}\,a_3, \label{cf3}\\
C_2&=&a_2^\prime=\frac{i\tau_2}{1-\rho_2\rho_{34}\,e^{2i\Phi_2}}\,a_2,\label{cf2}\\ 
C_1&=&a_1^\prime=\frac{i\tau_1}{1-\rho_1\rho_{234}\,e^{2i\Phi_1}}\,a_1\label{cf1}.
\end{eqnarray} 
\end{subequations}
For a detailed derivation of the expression for a \zmc{} refer for example to\cite{2mc}.
In Equation~(\ref{cf2}) and (\ref{cf1}) the abbreviations $\rho_{34}$ and $\rho_{234}$ are used to maintain the appearance of a simple \zmc{}. They stand for the reflection of the third cavity (behaving like an ordinary \zmc{}) (M$_{34}$) given by
\begin{equation}
\rho_{34}\left(\Phi_3\right)=\rho_3a_3+i\tau_3b_3=\frac{\rho_3-\rho_4e^{2i\Phi_3}}{1-\rho_3\rho_4e^{2i\Phi_3}}
\end{equation}
and the reflection of the \dmc{} containing M$_2$, M$_3$ and M$_4$ (labeled M$_{234}$ in the following) given by
\begin{equation}
\rho_{234}\left(\Phi_2,\Phi_3\right)=\frac{\rho_2-\rho_{34}e^{2i\Phi_2}}{1-\rho_2\rho_{34}e^{2i\Phi_2}}.
\end{equation}
Note that these complex expressions are frequency dependent values. Furthermore, the input field of the third resonator is $a_3=C_2e^{i\Phi_2}$ and that of the second one is $a_2=C_1e^{i\Phi_1}$. However, writing $C_k$ similar to the \zmc{} case and thinking of M$_{34}$ and M$_{234}$ as compound mirrors offers an easier understanding of these expressions.  

\subsection{Signal sidebands}
The basics of the response to gravitational waves are described for example in\cite{miz1} for a Michelson interferometer with ordinary two-mirror arm cavities. In this section we transfer these results to the \vmc{} case.\\
The normalized transfer function $\mathbf{G(\omega)}$ to the detection port for signal sidebands generated in the \vmc{} by gravitational waves (called `GW-transfer function' in the following) is composed of three parts corresponding to the three resonators formed by the \vmc{}. Each of the three parts is the sum of the \tf{} for the upper sidebands ($+\omega$) and the lower sidebands ($-\omega$).  
One obtains
\begin{equation}
\mathbf{G}\left(\omega\right)=\sum_{k=1,2,3}G_k\left(\omega\right)+G_k^*\left(-\omega\right)
\end{equation}
for the readout of the phase modulation and 
\begin{equation}
\mathbf{G}\left(\omega\right)=\sum_{k=1,2,3}-i\big(G_k\left(\omega\right)-G_k^*\left(-\omega\right)\big)
\end{equation}   
for the readout of the amplitude modulation. For an arbitrary operating point $\Phi_k^\text{op}$ the terms $G_k(\omega)$  have the form
\begin{equation}
\begin{split}
G_k(\omega)=\underbrace{\underbrace{C_k(\Phi_k^\text{op})}_{\text{carrier amplitude}}\times\underbrace{G^{\delta\phi}_k(\omega)}_{\text{modulation per round trip}}}_\text{sideband generation in the cavity}\dots\\ \times \underbrace{ C_k(\Phi_k^\text{op}+\omega L_k/c)}_{\text{\begin{tabular}{cc}sideband enhancement and\\ transfer to detection port\end{tabular}}}.
\end{split}\label{sbmod}
\end{equation}
Since sidebands generated by a traversing gravitational wave are impressed due to a modulation process their amplitudes are proportional to the carrier field. Accordingly the expression for the corresponding carrier field $C_k$ (first term of Equation~(\ref{sbmod})) is contained in $G_k(\omega)$. Note that here the input field $a_1$ is assumed to be unity because $G_k$ represents a normalized \tf{}. 

The expression describing the transfer from a gravitational wave to phase shift (modulation depth) is given by\cite{miz1} 
\begin{equation}
\mathop{X}_{h \rightarrow\phi}=\frac{\omega_0}{2}\frac{1-e^{i\omega l/c}}{-i\omega}\label{xmod}
\end{equation}
where $l$ is the optical path length. Each sideband impressed on the carrier $C_k$ by this weak phase modulation has the amplitude $C_kJ_1(X)=C_kX/2$. To obtain the magnitude of the sidebands generated per round trip in the respective cavity (second term in Eq.~(\ref{sbmod})), the amplitude attenuation (per round trip) of this cavity needs to be taken into account. In the case of the first and second cavity the reflecting mirror M$_\text{r}$ and thus the attenuation is frequency dependent. Accordingly, the first half and the second half round trip needs to be considered independently. One obtains
\begin{equation}
G_k^{\delta\phi}(\omega)=\left(\rho_k^\text{r}(\omega)e^{i\Phi_k}+\rho_k^\text{r}(\omega_0)e^{i\Phi_k^\text{op}}\right)\frac{\omega_0}{4}\frac{1-e^{i\frac{\omega L_k}{c}}}{-i\omega}
\end{equation}
where $\rho_k^\text{r}$ is the reflection of the respective reflecting mirror M$_\text{r}$ (i.e. in the case of the first cavity $\rho_k^\text{r}$ is $\rho_{234}$). The first term in the parentheses describes the attenuation of the sidebands generated on the first half round trip. These sidebands are reflected at M$_{234}$ with $\rho_{234}(\omega)$. The phase $\Phi_k=\Phi_k^\text{op}+\omega L_k/c$ accounts for the delay incurring on the second half round trip. The second term in the parentheses describes the sidebands generated on the second half round trip. The amplitude of these sidebands is proportional to the reflected carrier field. Thus in this case the attenuation $\rho_k^\text{r}(\omega_0)$ is that of the carrier amplitude. Here the phase factor $e^{i\Phi_k^\text{op}}$ describes the carriers phase delay incurred on the first half round trip. 

In addition the phase relation between the signal sidebands induced in different cavities needs to be included. Thus, the geometric layout of the resonators (refer to Figure~\ref{ifo}) needs to be taken into account as well. If the \vmc{} is not folded, signal sidebands of gravitational waves are impressed in common mode. But if for example the first and third resonator are orientated orthogonally to each other, the corresponding sidebands are impressed differentially (due to the quadrupole nature of gravitational waves). Thus, in the differential mode $G_1^{\delta\phi}$ and $G_3^{\delta\phi}$ have different signs.  

Whereas the first two terms of Eq.~(\ref{sbmod}) principally describe the sideband generation in the respective cavities, the frequency response is mainly determined by the third term in Eq.~(\ref{sbmod}). The frequency dependent enhancement of the sidebands inside the cavities and the transfer out of the cavities to the detection port can be obtained from $C_k$ by substituting
\begin{equation}
C_k\left(\Phi_k\right)\rightarrow C_k\left(\Phi_k^\text{op}+\frac{\omega L_k}{c}\right)\label{term2}     
\end{equation}
where $\omega L_k/c$ describes the phase delay incurring while propagating over the length $L_k$. This becomes clear considering for example the first cavity. The expression $C_1$ describes the enhancement of the amplitude  $a_1$ injected at M$_1$ into the first cavity (refer to Figure~\ref{notation}). The signal sidebands are generated inside the cavities, get resonantly enhanced, and are then transmitted through M$_1$ to the detection port. Hence we do not need to multiply the sideband amplitudes by $\tau_1$ upon injection but during the  extraction (transmission through M$_1$) of the cavity and Eqs.~(\ref{cf3})-(\ref{cf1}) are valid not only for the externally injected carrier but also for the internally generated sidebands. 

The  shot noise limited sensitivity is given by the noise to signal ratio of the signal $\mathbf{G}\left(\omega\right)$ and the phase fluctuation equivalent to shot noise\cite{winkblair}
\begin{equation}
\tilde{\delta\phi}=\sqrt{\frac{2\hbar\omega_0}{P_0}} 
\end{equation}
leading to
\begin{equation}
\tilde h (\omega)=\frac{1}{2\left|\mathbf{G}(\omega)\right|}\sqrt{\frac{2\hbar\omega_0}{P_0}}\label{strainsens},
\end{equation}
where $P_0$ is the power of the incident carrier light. Note, that $\mathbf{G(\omega)}$ is calculated for a single interferometer arm. This results in an additional factor of 1/2 for the sensitivity of the whole interferometer\cite{miz3}.  
Since this sensitivity depends on eleven parameters (3 macroscopic lengths, 3 tunings, 4 reflectivities and the geometric layout), a qualitative and intuitive understanding of the resonance feature is necessary to find some basic assumptions and boundary conditions which allow a systematic analysis of the configuration.   

\section{\label{ropr}Reduction of parameter range}
\begin{figure*}
\includegraphics[scale=0.555]{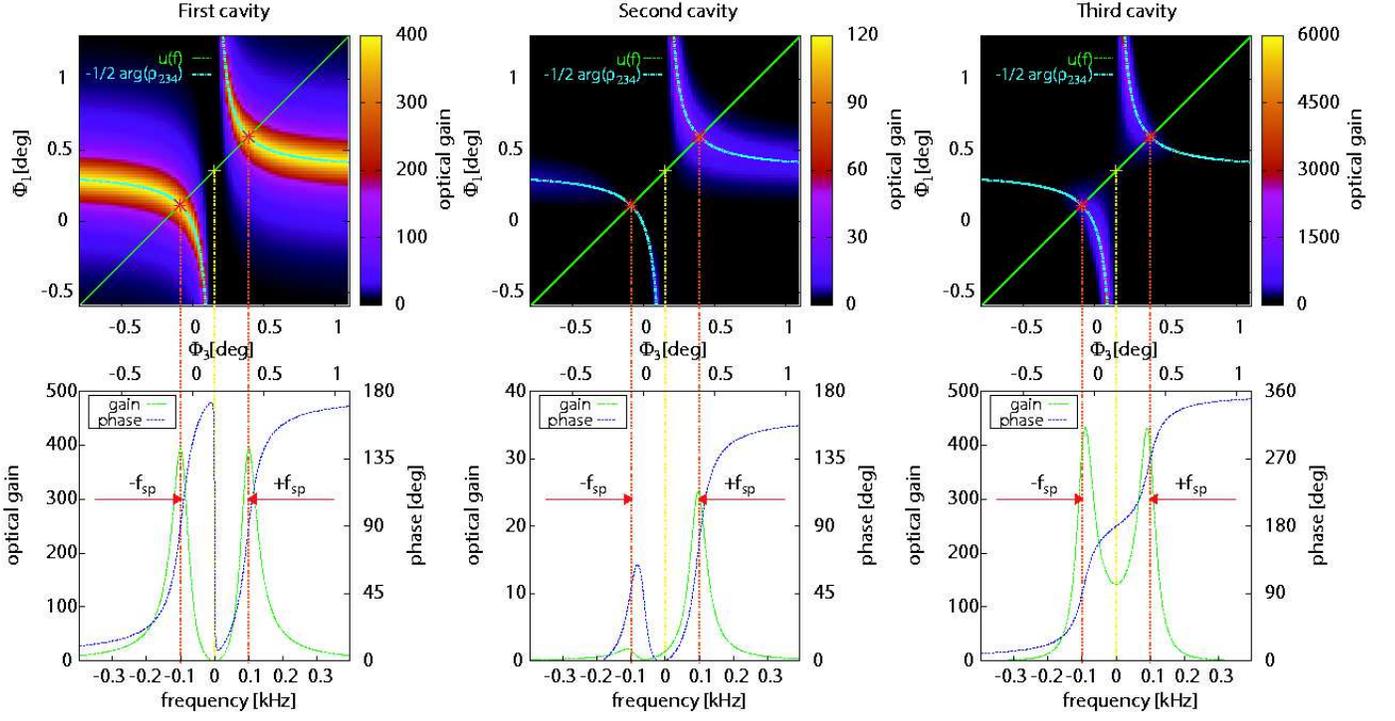}
\caption{\label{ident}(Color online) For easier readability refer to the colored online version. The upper three graphs show  the optical gains $\left|C_k\right|^2$  in dependence of $\Phi_3$ and $\Phi_1$. The tuning of the second cavity is fixed. The cyan curve illustrates that the loci of maximum optical gain (resonance) can be related to the phase shift $\arg(\rho_{234})$ in reflection of M$_{234}$ (refer to Eq.~(\ref{reso})). Note that these local maxima are determined in dependence of $\Phi_3$ and $\Phi_2$. Accordingly, the loci correspond to maxima in cross sections (cuts) parallel to the $\Phi_1$-axis. The cut through the $\Phi_3$-$\Phi_1$-plane corresponding to the frequency response of each cavity is shown with the green lines ($u(f)$). Here $L_1$ and $L_3$ are 2\,km each. The point symmetry of the resonance condition is illustrated with the yellow crosses. The red crosses mark the intersections of $u(f)$ with the loci of maximum optical gains (obtained from solving the RHS of Eq.~(\ref{reso}) for Eq.~(\ref{cut})).\\ 
The lower graphs show the frequency response of the optical gains (green curves) and the phasing (blue curves)  along the green line ($u(f)$) in the upper graphs. The frequency scale is calculated as $f=(\Phi_3^\text{sym}-\Phi_3)c/L_3$. Thus, $f=0\,\text{Hz}$ is related to the symmetry point. The frequency splitting $\omega_\text{sp}=2\pi f_\text{sp}$ is defined with respect to this symmetry point. The connections to the upper graphs demonstrates that the responses of the first and third resonator are symmetric (yellow line) and furthermore that resonances in the frequency response can be pre-estimated by the intersection of $u(f)$ with the loci of resonance.}
\end{figure*}

\subsection{Basic assumptions}
Since the \vmc{} is meant as an alternative to RSE-topologies, we will derive some basic boundary conditions for the comparison with the Advanced LIGO optical configuration\cite{ligo}. 
In consequence, the maximum interferometer arm length is set to 4\,km. The calculated strain sensitivities are related to an input power of 125\,W. The transmission of the end mirror M$_4$ is set to 50\,ppm. In addition, to avoid thermal effects in the optical substrates embedded in the second cavity, the power in this cavity needs to be relatively small. Thus, signal sidebands induced in this cavity are small compared to those induced in the first and third one. Accordingly, the length $L_2$ does not need to be long to enhance the sensitivity for gravitational waves. It can be shown that even the shape of the GW-\tf s is not significantly affected by $L_2$. Thus, $L_2$ is arbitrarily set to 10\,\text{m} for all calculations. Hence the second cavity can be understood as an etalon whose transmission $T_\text{c}$ determines the coupling of the first and third cavity and thus the frequency splitting of the corresponding resonance doublet.
This is one key feature of the topology. The \vmc{} behaves like a \dmc{} with variable coupling. The properties of the second cavity solely determine the frequency splitting.

\subsection{Resonance feature}  
The transmission of an ordinary \zmc{} becomes maximum on resonance. In analogy to this, the resonance condition of the \vmc{} corresponds to local maxima in the transmitted field. Concerning a \zmc{}, the incident light resonates if the phase shift per round trip equals 0 mod(2$\pi$). Here the phase shift comes from the phase delay $2\omega L/c$ incurred while traveling twice the macroscopic length $L$. In the case of the \vmc{}, the phasing in the first cavity is composed of the corresponding delay $2\omega L_1/c$ and the phase shift $\arg\left(\rho_{234}\right)$ in reflection of M$_{234}$. For given reflectivities and tunings $\Phi_3$ and $\Phi_2$, the first cavity can be understood as a \zmc{} with M$_{234}$ as the end mirror giving an additional phase shift. Thus the resonance condition (leading to a maximum light field) in the first cavity can be determined according to
\begin{equation}
2\underbrace{\frac{\omega L}{c}}_{\Phi_1}=-\arg\left(\rho_{234}\left(\Phi_2,\Phi_3\right)\right)\label{reso}
\end{equation}
(shown by the cyan line in the upper graphs of Figure~\ref{ident}). Since the enhancements of the second cavity ($C_2$) and the third cavity ($C_3$) are constant if $\Phi_2$ and $\Phi_3$ are fixed (refer to Eqs.~(\ref{cf3}) and (\ref{cf2})), a maximum light field in the first cavity leads to maximum light fields in the second and third cavity. This relation drastically reduces  the parameter range for the resonances of the \vmc{}.

\subsection{Power in substrates} 
Another boundary condition can be derived from the point symmetry in the loci of maximum optical gain. Concerning the pattern of the internal fields in the $\Phi_3$-$\Phi_1$-plane, those of the first and third resonator are point symmetric whereas that of the second one is not (refer to Figure~\ref{ident}). Considering the absolute values of the carrier fields $C_k$ (Eqs.~(\ref{cf3})-(\ref{cf1})) reveals that the carrier fields are point symmetric in the tunings $\Phi_k$. Due to the fact that all tunings in $C_k$ appear with the factor $i$ the relation 
\begin{equation}
\left|C_k\left(\Phi_k\right)\right|=\left|C_k^*\left(\Phi_k\right)\right|=\left|C_k\left(-\Phi_k\right)\right| \label{sym}
\end{equation}
is valid explaining a point symmetry around the origin $(\Phi_k=0)$. Since in general $\Phi_2$ is not zero but tuned ($\Phi_2^\text{c}$) to give a certain frequency splitting, the relation of Eq.~(\ref{sym}) is not fulfilled. In this case 
\begin{equation}
\left|C_k(\Phi_1,\Phi_2^\text{c},\Phi_3)\right|\neq\left|C_k(-\Phi_1,\Phi_2^\text{c},-\Phi_3)\right| 
\end{equation}
showing that $\left|C_2\right|^2$ is not point symmetric in the $\Phi_3$-$\Phi_1$-plane. The symmetry still existent in the pattern of $\left|C_1\right|^2$ and $\left|C_3\right|^2$ comes from the point symmetric phase shift $\arg(\rho_{234})$ in reflection around the resonance in M$_{234}$. Here M$_{234}$ is considered as a \zmc{} with M$_{23}$ as the coupling mirror. Accordingly the phase shift in reflection of M$_{234}$ must correspond to the point symmetric \zmc{} case. From this phase shift the coordinates of the symmetry point can be determined in analogy to Eq.~(\ref{reso}) as
\begin{eqnarray}
\Phi_3^\text{sym}&=&-\frac{1}{2}\arg\left(\rho_{32}(\Phi_2^\text{c})\right)\hspace{10pt}\text{and}\label{xsym}\\
\Phi_1^\text{sym}&=&-\frac{1}{2}\arg\left(\rho_{234}(\Phi_2^\text{c},\Phi_3^\text{sym})\right).\label{ysym}
\end{eqnarray}
Note, that in Eq.~(\ref{xsym}) $\Phi_3^\text{sym}$ is determined in reflection of the second cavity (etalon). This is possible because of the reciprocity concerning the resonance in M$_{234}$. Here reciprocity means that the transmission and thus the resonance condition of M$_{234}$ must not differ if the light is injected at M$_4$ instead of M$_2$.

Then, to fulfill the condition of low power in the second resonator, only those operating points lying on the lower left resonance branch are suitable (refer to the optical gain of the second cavity shown in the middle of Figure~\ref{ident}). Since the pattern of the internal carrier fields are only determined by the reflectivities of the mirrors, appropriate operating points with respect to the power buildup can be investigated independently from the macroscopic cavity lengths $L_k$. The choice of the lengths concerns only the GW-\tf{}.

\subsection{Frequency response}
The frequency response of the \vmc{} is represented by a cut through the $\Phi_3$-$\Phi_1$-plane (upper graphs in Figure~\ref{ident}) along an oblique line given by
\begin{equation}
u(\omega)=\frac{L_1}{L_3} \left(\Phi_3-\Phi^\textrm{op}_3\right)+\Phi^\textrm{op}_1,\label{cut}
\end{equation}
where $\Phi_3^\textrm{op}$ and $\Phi_1^\textrm{op}$ corresponds to the operating point of the carrier light. The slope of this line is determined by the ratio $L_1/L_3$ because the tunings change with the frequency according to $d\Phi_k/d\omega\propto L_k$.  Thus, resonances in the transfer function are indicated by the intersections of $u(\omega)$ with the loci of maximum optical gain given by Eq.~(\ref{reso}). The frequency splitting  $\omega_\text{sp}$ is given by the coordinates of these intersection points and is counted from the symmetry point (refer to the lower graphs of Figure~\ref{ident}). The loci of maximum optical gain are determined in dependence of given $\Phi_3$ (and $\Phi_2^\text{c}$). Thus, they are related to maxima along cuts parallel to the $\Phi_1$-axis. Accordingly these maxima are not necessarily local maxima along the line $u(\omega)$. However, it is useful to pre-estimate resonances in the \tf s considering these intersections (see below). As the transmission $T_\text{c}$ of the second cavity determines the coupling between the first and third resonator and thus the frequency splitting $\omega_\text{sp}$  an expression can be derived giving $T_\text{c}$ as a function of this frequency splitting. For this purpose we consider a \dmc{} with equal length $L_1^\prime$ and $L_2^\prime$ and an end mirror with $\rho_3^\prime=\rho_4$. Here the transmission (also $T_\text{c}$) of the second mirror determines the frequency splitting. Solving the intersection $(-1/2)\arg(\rho_{23}^\prime(\omega_\text{sp}L_2^\prime/c))=(L_1^\prime/L_2^\prime)\omega_\text{sp}L_2^\prime/c$ for $\rho_2^\prime$ gives
\begin{equation}
T_\text{c}=1-(\rho_2^\prime)^2=1-\underbrace{\frac{4\,\cos^2(2\frac{\omega_\textrm{sp}L_2^\prime}{c})\rho_4^2}{\left(1+\rho_4^2\right)^2}}_{R_\text{c}}.\label{tc}
\end{equation} 
Note that the resonance frequencies of the single resonators depend on their macroscopic lengths according to $FSR_k=c/2L_k$. Thus, the frequency splitting $\omega_\text{sp}$ of the coupled resonators  is related to the lengths $L_1$ and $L_3$, too. In the following explicit values of $\omega_\text{sp}$ always refer to $L_1=L_3=2\,\text{km}$ and to the symmetry point.
For fixed reflectivities $\rho_2$ and $\rho_3$ the tuning $\Phi_2^\textrm{c}$ leading to the desired $T_\textrm{c}$  is given by
\begin{equation}
\Phi_2^\text{c}=\frac{1}{2}\arccos\left(\frac{R_\text{c}-\rho_2^2-\rho_3^2+R_\text{c}\rho_2^2\rho_3^2}{2\rho_2\rho_3T_\text{c}}\right).\label{phi2c}
\end{equation}
Since the effective transmission $T_\textrm{c}$ of the second cavity is the decisive parameter for the frequency splitting, $\rho_2$ and $\rho_3$ initially can be chosen almost arbitrarily. Only the coordinates of the symmetry point depend on $\rho_2$ and $\rho_3$. The pattern of $\left|C_1\right|^2$ and $\left|C_3\right|^2$ related to this symmetry point stay the same. Later the optimal values are determinable by the  boundary condition of low power in the second cavity. Accordingly, for a certain $T_\textrm{c}$ the enhancement of the carrier light only depends on $\Phi_1$, $\Phi_3$ and $\tau_1$ ($\tau_4$ is 50\,ppm). The GW-\tf{} additionally depends on $L_1$ and $L_3$. Thus, the parameter space for the sensitivity of the investigated topology is reduced to a six dimensional one. Additionally, $\Phi_2$ can be chosen from Eq.~(\ref{phi2c}) to give a certain frequency splitting. Then, the symmetry point from Eqs.~(\ref{xsym}) and (\ref{ysym}) serves as starting point for $\Phi_3$ and $\Phi_1$ leading to a reduction of the parameter range for the tunings. 
\section{\label{study}Parameter studies}
\begin{figure*}
\psfrag{sensitivity  [1/texHz]}{\fontfamily{Helvetica}\fontsize{9}{4}{sensitivity $[1/\sqrt{\text{Hz}}]$}}
\psfrag{frequency [Hz]}{\fontfamily{Helvetica}\fontsize{9}{7}{frequency [Hz]}}
\includegraphics[scale=0.7]{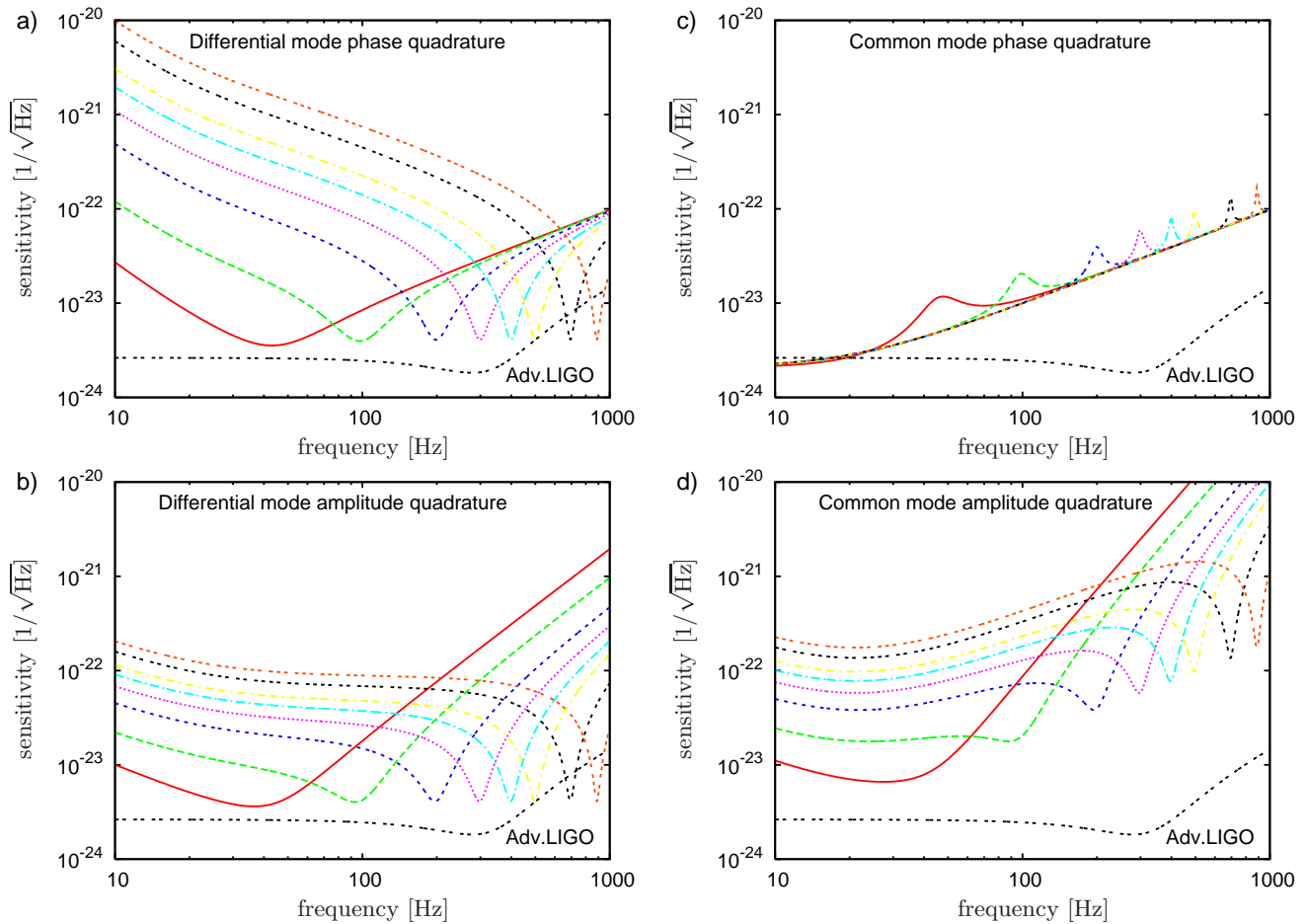}
\caption{\label{Phi2dep}(Color online) Sensitivity of the \vmc{} for various frequency splittings $f_\text{sp}=\omega_\text{sp}/2\pi$ induced by the tuning $\Phi_2^\text{c}$. The graphs a) and b) show the sensitivities in phase and amplitude quadrature for a folded geometry (differential mode). Graphs c) and d) correspond to the unfolded setup (common mode). It can be seen that the best sensitivities can be achieved in the differential mode.}
\end{figure*}  
\subsection{Special cases}
The analysis of the resonance feature in Section~\ref{ropr} revealed that the resonance of the carrier light does not depend on the macroscopic lengths $L_k$, because the resonance is determined by the microscopic lengths (tunings). Furthermore, it was stated that the sensitivity for gravitational waves is proportional to the carrier amplitudes in the cavities. Thus, operating points yielding satisfying sensitivities are restricted to states where the carrier is on resonance (or close to resonance). For given reflectivities these states are determined by Eq.~(\ref{reso}). First we investigate the power buildup (optical gain) in the first and third cavity in dependence of the frequency splitting  induced by the tuning $\Phi_2^\text{c}$ of the second cavity. Here, the reflectivities $\rho_2^2$ and $\rho_3^2$ were set to 0.999 each and $\rho_1^2$ was set to 0.993. The frequency splitting $f_\text{sp}=\omega_\text{sp}/2\pi$ was investigated in a range from 50\,Hz to 1\,kHz corresponding to the detection band of terrestrial gravitational-wave detectors. Furthermore, the respective operating points were chosen as the intersection point (located on the lower left resonance branch) of $u(\omega)$ with the loci of maximum optical gain. Note, that the carrier is resonant on these states (refer to Figure~\ref{ident}). Then the upper sidebands are expected to be resonant at frequencies $\omega=2\omega_\text{sp}$. The corresponding tuning $\Phi_2^\text{c}$ was determined by Equations~(\ref{tc}) and (\ref{phi2c}). The cavity lengths $L_1$ and $L_3$ were set to 2\,km each.  This investigation revealed that both the optical gain and the resonances' bandwidth in these special states are independent of the frequency splitting in the investigated region. Figure~\ref{Phi2dep} shows the shot noise limited sensitivities for various tunings $\Phi_2^\text{c}$.
 
\begin{figure}[h!] 
\psfrag{sensitivity [1/texHz]}{\fontfamily{Helvetica}\fontsize{9}{4}{sensitivity $[1/\sqrt{\text{Hz}}]$}}
\psfrag{frequency [Hz]}{\fontfamily{Helvetica}\fontsize{9}{7}{frequency [Hz]}}
\includegraphics[scale=0.67]{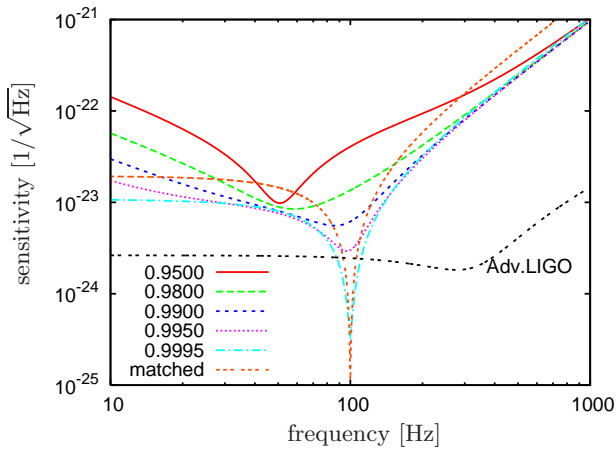}
\caption{\label{R1dep}(Color online) Sensitivity in the amplitude quadrature for the differential mode with different parameters $\rho_1^2$. Here the tuning $\Phi_2^\text{c}$ corresponds to a splitting of 50\,Hz.}
\end{figure}

Since for these states the tunings and the cavity lengths were determined with respect to certain frequency splittings and resonances in the frequency response, $\rho_1$ is the only remaining free parameter affecting the optical gains. Accordingly, these states are investigated for the dependence on $\rho_1$. The results are shown in Figure~\ref{R1dep}. It can be seen that the accuracy of the estimation of resonances in the response depends on $\rho_1$. This is because on the one hand the intersection points indicating these resonances correspond to local maxima along $\Phi_1$ and not along $u(\omega)$. On the other hand these intersection points were determined independent of $\rho_1$. Considering an extreme case clarifies that the resonances in the frequency response change with $\rho_1$. If $\rho_1\approx0$ is assumed, the influence of the first cavity is negligible. In this case, the resonance feature is dominated by M$_{234}$ whose resonance corresponds to the symmetry point. Accordingly, for a low $\rho_1$ the frequency response becomes  maximum on the symmetry point explaining the peak sensitivity around 50\,Hz for $\rho_1=0.95$ (solid curve in Figure~\ref{R1dep}). Furthermore it can be seen, that the impedance matched case optimally exploiting the laser input field yields a better sensitivity compared to Advanced LIGO only in a very narrow region.
 
\begin{figure}
\psfrag{sensitivity [1/texHz]}{\fontfamily{Helvetica}\fontsize{9}{4}{sensitivity $[1/\sqrt{\text{Hz}}]$}}
\psfrag{frequency [Hz]}{\fontfamily{Helvetica}\fontsize{9}{7}{frequency [Hz]}}
\includegraphics[scale=0.65]{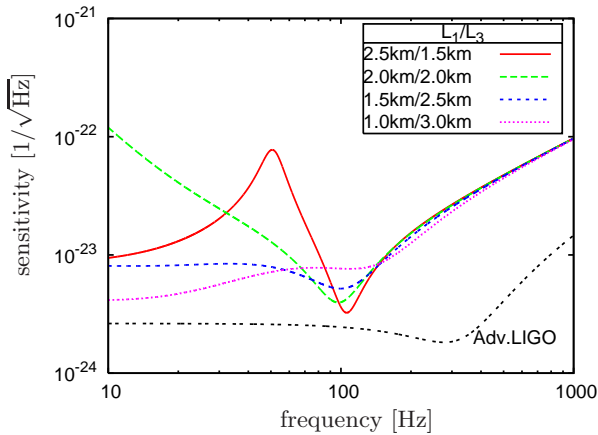}
\caption{\label{Ldep}(Color online) Sensitivity in the phase quadrature of the differential mode with different ratios of $L_1/L_3$.}
\end{figure}

Figure~\ref{Ldep} shows the sensitivity for different ratios $L_1/L_3$. This ratio determines the slope of the oblique line $u(f)$ in Figure~\ref{ident} representing the frequency response. Accordingly, the intersection points of this line with the loci of resonance vary with this ratio.
Also the magnitude of the signal sidebands induced in a cavity scale with its length (refer to Eq.~\ref{xmod})). Hence if $L_1$ is greater than $L_3$ the GW-transfer function is  dominated by the sidebands induced in the first cavity (if the powers in the first and third cavity are similar).
The  optical gain $\left|C_1\right|^2$ is very low around the symmetry point as it can be seen in Figure~\ref{ident}. This low optical gain corresponds to the bad sensitivity around 50\,Hz for $L_1/L_3=2.5\,\text{km}/1.5\,\text{km}$ (solid curve in Figure~\ref{Ldep}). This fact indicates, that satisfying sensitivities in a broad frequency range are only achievable for configurations with $L_1/L_3\leq1$. Also configurations with very short $L_1$ (or $L_3$) yield no satisfying sensitivities. In these cases, the \vmc{} behaves like an ordinary \zmc{} with only one resonance frequency given at $c/2L_3$ (or $c/2L_1$).      

All other non special cases  were investigated with a numerical code developed for this purpose. With this code, the parameters were systematically varied considering all relations and dependencies obtained in Section~\ref{ropr}. The frequency splitting $f_\text{sp}$ was varied between 50\,Hz and 1000\,Hz. The tunings $\Phi_3$ and $\Phi_1$ were scanned around the lower left resonance branch. Various ratios of $L_1/L_3$ were investigated. As already expected from the investigation of the special cases described above, there were no impedance matched configurations found having sensitivities comparable in a wide frequency range to the Advanced LIGO optical configuration. Thus, also for the \vmc{} the Power Recycling technique comes in consideration to broaden the bandwidth for signal sidebands.  

\subsection{Exemplary configurations}
\begin{figure}
\psfrag{sensitivity [1/texHz]}{\fontfamily{Helvetica}\fontsize{9}{4}{sensitivity $[1/\sqrt{\text{Hz}}]$}}
\psfrag{frequency [Hz]}{\fontfamily{Helvetica}\fontsize{9}{7}{frequency [Hz]}}
\includegraphics[scale=0.65]{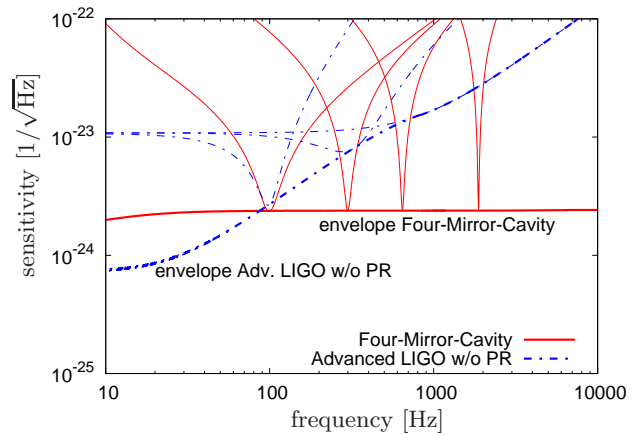}
\caption{\label{examp1}(Color online) Comparison of the envelopes of the tunable peak sensitivity in the phase quadrature. Exemplary sensitivity curves with different tunings but fixed reflectivities are shown.}
\end{figure}

Here we present the properties of one exemplary configuration which yields sensitivities typical for a \vmc{} enhanced Michelson interferometer. Figure~\ref{examp1} shows the envelope of the tunable peak sensitivity and exemplary sensitivity curves for different frequency splittings, whereas the reflectivities are fixed. The reflectivities $\rho_2=\rho_3=0.999$ were chosen to ensure low powers in the second cavity embedding optical substrates. The reflectivity $\rho_1=0.996$ was chosen as compromise between a high peak sensitivity and a broad bandwidth. With this setup, the power in the first and third cavity are identical (approximately 61\,kW with 125\,W input power) and independent of the frequency splitting. The power in the second cavity is $P_2\approx16$\,W. Thus, the boundary condition of low powers in optical substrates can be fulfilled with remarkable low powers in the second cavity. The sensitivity of this configuration is as good as the Advanced LIGO ones, if identical powers at the beam splitter are assumed. It should be mentioned, that the \vmc{} can be tuned to high frequencies with a constant peak sensitivity in contrast to RSE-topologies. Also the detection bandwidth is not limited by the arm cavities' finesse (refer to the envelopes shown in figure~\ref{examp1}), because the effective finesse is adjustable by the transmission of the second cavity. This fact also implies that the intra cavity losses in the second cavity caused by the AR-coatings and absorption in the optics' substrates are not limiting the performance of the \vmc{}.  

\section{Conclusion}
We derived the expressions describing the shot noise limited sensitivity of a \vmc{} enhanced Michelson interferometer. The detailed analysis of the resonance feature using expressions similar to the ordinary \zmc{} case offers a qualitative and intuitive understanding of this complex configuration. With this understanding it was possible to systematically investigate the configuration for shot noise limited sensitivities throughout the whole 11-dimensional parameter space. Despite the huge parameter space giving a variety of possibilities to adapt the shape of the sensitivity curve to the requirements, contrary to expectations there are no parameter configurations optimally exploiting both the frequency splitting and the laser input field yielding better sensitivities compared to Advanced LIGO in a wide frequency range. Only if identical powers at the beam splitter are assumed, the \vmc{} provides sensitivities comparable to the Advanced LIGO ones. However, these powers at the beam splitter would still be restricted by the problem of thermal lensing. Thus, our results confirm the choice of RSE-topologies for the second generation of interferometric gravitational detectors.    

\bibliography{thuering-4mc.bib}

\begin{thebibliography}{19}
\expandafter\ifx\csname natexlab\endcsname\relax\def\natexlab#1{#1}\fi
\expandafter\ifx\csname bibnamefont\endcsname\relax
  \def\bibnamefont#1{#1}\fi
\expandafter\ifx\csname bibfnamefont\endcsname\relax
  \def\bibfnamefont#1{#1}\fi
\expandafter\ifx\csname citenamefont\endcsname\relax
  \def\citenamefont#1{#1}\fi
\expandafter\ifx\csname url\endcsname\relax
  \def\url#1{\texttt{#1}}\fi
\expandafter\ifx\csname urlprefix\endcsname\relax\def\urlprefix{URL }\fi
\providecommand{\bibinfo}[2]{#2}
\providecommand{\eprint}[2][]{\url{#2}}

\bibitem[{\citenamefont{Mizuno et~al.}(1993)\citenamefont{Mizuno, Strain,
  Nelson, Chen, Schilling, R{\"u}diger, Winkler, and Danzmann}}]{miz2}
\bibinfo{author}{\bibfnamefont{J.}~\bibnamefont{Mizuno}},
  \bibinfo{author}{\bibfnamefont{K.~A.} \bibnamefont{Strain}},
  \bibinfo{author}{\bibfnamefont{P.~G.} \bibnamefont{Nelson}},
  \bibinfo{author}{\bibfnamefont{J.~M.} \bibnamefont{Chen}},
  \bibinfo{author}{\bibfnamefont{R.}~\bibnamefont{Schilling}},
  \bibinfo{author}{\bibfnamefont{A.}~\bibnamefont{R{\"u}diger}},
  \bibinfo{author}{\bibfnamefont{W.}~\bibnamefont{Winkler}}, \bibnamefont{and}
  \bibinfo{author}{\bibfnamefont{K.}~\bibnamefont{Danzmann}},
  \bibinfo{journal}{Phys. Lett A} \textbf{\bibinfo{volume}{175}},
  \bibinfo{pages}{273} (\bibinfo{year}{1993}).

\bibitem[{\citenamefont{Meers}(1988)}]{meers}
\bibinfo{author}{\bibfnamefont{B.}~\bibnamefont{Meers}},
  \bibinfo{journal}{Phys. Rev D} \textbf{\bibinfo{volume}{38}},
  \bibinfo{pages}{2317} (\bibinfo{year}{1988}).

\bibitem[{\citenamefont{Weinstein}(2002)}]{ligo}
\bibinfo{author}{\bibfnamefont{A.}~\bibnamefont{Weinstein}},
  \bibinfo{journal}{Class. Quantum Grav.} \textbf{\bibinfo{volume}{19}},
  \bibinfo{pages}{1575} (\bibinfo{year}{2002}).

\bibitem[{\citenamefont{Drever et~al.}(1983)\citenamefont{Drever, Hough,
  Munley, Lee, Spero, Whitecomb, Ward, Ford, Hereld, Robertson
  et~al.}}]{drever-pr}
\bibinfo{author}{\bibfnamefont{R.~W.~P.} \bibnamefont{Drever}},
  \bibinfo{author}{\bibfnamefont{J.}~\bibnamefont{Hough}},
  \bibinfo{author}{\bibfnamefont{A.~J.} \bibnamefont{Munley}},
  \bibinfo{author}{\bibfnamefont{S.-A.} \bibnamefont{Lee}},
  \bibinfo{author}{\bibfnamefont{R.}~\bibnamefont{Spero}},
  \bibinfo{author}{\bibfnamefont{S.~E.} \bibnamefont{Whitecomb}},
  \bibinfo{author}{\bibfnamefont{H.}~\bibnamefont{Ward}},
  \bibinfo{author}{\bibfnamefont{G.~M.} \bibnamefont{Ford}},
  \bibinfo{author}{\bibfnamefont{M.}~\bibnamefont{Hereld}},
  \bibinfo{author}{\bibfnamefont{N.}~\bibnamefont{Robertson}},
  \bibnamefont{et~al.}, \emph{\bibinfo{title}{Quantum Optics, Experimental
  Gravitation and Measurement Theory}} (\bibinfo{publisher}{Plenum Press},
  \bibinfo{address}{New York}, \bibinfo{year}{1983}), \bibinfo{note}{p.321}.

\bibitem[{\citenamefont{Winkler et~al.}(1991)\citenamefont{Winkler, Danzmann,
  R{\"u}diger, and Schilling}}]{wink}
\bibinfo{author}{\bibfnamefont{W.}~\bibnamefont{Winkler}},
  \bibinfo{author}{\bibfnamefont{K.}~\bibnamefont{Danzmann}},
  \bibinfo{author}{\bibfnamefont{A.}~\bibnamefont{R{\"u}diger}},
  \bibnamefont{and}
  \bibinfo{author}{\bibfnamefont{R.}~\bibnamefont{Schilling}},
  \bibinfo{journal}{Phys. Rev A} \textbf{\bibinfo{volume}{44}},
  \bibinfo{pages}{7022} (\bibinfo{year}{1991}).

\bibitem[{\citenamefont{Strain et~al.}(1991)\citenamefont{Strain, Danzmann,
  Mizuno, Nelson, R{\"u}diger, Schilling, and Winkler}}]{strain}
\bibinfo{author}{\bibfnamefont{K.~A.} \bibnamefont{Strain}},
  \bibinfo{author}{\bibfnamefont{K.}~\bibnamefont{Danzmann}},
  \bibinfo{author}{\bibfnamefont{J.}~\bibnamefont{Mizuno}},
  \bibinfo{author}{\bibfnamefont{P.~G.} \bibnamefont{Nelson}},
  \bibinfo{author}{\bibfnamefont{A.}~\bibnamefont{R{\"u}diger}},
  \bibinfo{author}{\bibfnamefont{R.}~\bibnamefont{Schilling}},
  \bibnamefont{and} \bibinfo{author}{\bibfnamefont{W.}~\bibnamefont{Winkler}},
  \bibinfo{journal}{Phys. Lett. A} \textbf{\bibinfo{volume}{194}},
  \bibinfo{pages}{124} (\bibinfo{year}{1991}).

\bibitem[{\citenamefont{McClelland et~al.}(1999)\citenamefont{McClelland, Camp,
  Mason, Kells, and Whitcomb}}]{McClel}
\bibinfo{author}{\bibfnamefont{D.~E.} \bibnamefont{McClelland}},
  \bibinfo{author}{\bibfnamefont{J.~B.} \bibnamefont{Camp}},
  \bibinfo{author}{\bibfnamefont{J.}~\bibnamefont{Mason}},
  \bibinfo{author}{\bibfnamefont{W.}~\bibnamefont{Kells}}, \bibnamefont{and}
  \bibinfo{author}{\bibfnamefont{E.}~\bibnamefont{Whitcomb}},
  \bibinfo{journal}{Opt. Lett.} \textbf{\bibinfo{volume}{24}},
  \bibinfo{pages}{1014} (\bibinfo{year}{1999}).

\bibitem[{\citenamefont{Tomaru et~al.}(2002)}]{saph}
\bibinfo{author}{\bibfnamefont{T.}~\bibnamefont{Tomaru}} \bibnamefont{et~al.},
  \bibinfo{journal}{Class. Quantum Grav.} \textbf{\bibinfo{volume}{19}},
  \bibinfo{pages}{2045} (\bibinfo{year}{2002}).

\bibitem[{\citenamefont{L{\"u}ck et~al.}(2000)\citenamefont{L{\"u}ck,
  M{\"u}ller, Aufmuth, and Danzmann}}]{Lueck}
\bibinfo{author}{\bibfnamefont{H.}~\bibnamefont{L{\"u}ck}},
  \bibinfo{author}{\bibfnamefont{K.-O.} \bibnamefont{M{\"u}ller}},
  \bibinfo{author}{\bibfnamefont{P.}~\bibnamefont{Aufmuth}}, \bibnamefont{and}
  \bibinfo{author}{\bibfnamefont{K.}~\bibnamefont{Danzmann}},
  \bibinfo{journal}{Opt. Comm} \textbf{\bibinfo{volume}{175}},
  \bibinfo{pages}{275} (\bibinfo{year}{2000}).

\bibitem[{\citenamefont{Lawrence et~al.}(2002)\citenamefont{Lawrence, Zucker,
  Fritschel, Marfuta, and Schumaker}}]{law1}
\bibinfo{author}{\bibfnamefont{R.}~\bibnamefont{Lawrence}},
  \bibinfo{author}{\bibfnamefont{M.}~\bibnamefont{Zucker}},
  \bibinfo{author}{\bibfnamefont{P.}~\bibnamefont{Fritschel}},
  \bibinfo{author}{\bibfnamefont{P.}~\bibnamefont{Marfuta}}, \bibnamefont{and}
  \bibinfo{author}{\bibfnamefont{D.}~\bibnamefont{Schumaker}},
  \bibinfo{journal}{Class. Quantum Grav.} \textbf{\bibinfo{volume}{19}},
  \bibinfo{pages}{1803} (\bibinfo{year}{2002}).

\bibitem[{\citenamefont{Lawrence et~al.}(2004)\citenamefont{Lawrence, Ottaway,
  Zucker, and Fritschel}}]{law2}
\bibinfo{author}{\bibfnamefont{R.}~\bibnamefont{Lawrence}},
  \bibinfo{author}{\bibfnamefont{D.}~\bibnamefont{Ottaway}},
  \bibinfo{author}{\bibfnamefont{M.}~\bibnamefont{Zucker}}, \bibnamefont{and}
  \bibinfo{author}{\bibfnamefont{P.}~\bibnamefont{Fritschel}},
  \bibinfo{journal}{Opt. Lett.} \textbf{\bibinfo{volume}{29}},
  \bibinfo{pages}{2635} (\bibinfo{year}{2004}).

\bibitem[{\citenamefont{Degallaix et~al.}(2004)\citenamefont{Degallaix, Zhao,
  Ju, and Blair}}]{blair}
\bibinfo{author}{\bibfnamefont{J.}~\bibnamefont{Degallaix}},
  \bibinfo{author}{\bibfnamefont{C.}~\bibnamefont{Zhao}},
  \bibinfo{author}{\bibfnamefont{L.}~\bibnamefont{Ju}}, \bibnamefont{and}
  \bibinfo{author}{\bibfnamefont{D.}~\bibnamefont{Blair}},
  \bibinfo{journal}{Class. Quantum Grav.} \textbf{\bibinfo{volume}{21}},
  \bibinfo{pages}{903} (\bibinfo{year}{2004}).

\bibitem[{\citenamefont{Drever}(1995)}]{drever}
\bibinfo{author}{\bibfnamefont{R.~W.~P.} \bibnamefont{Drever}},
  \emph{\bibinfo{title}{Proceedings of the Seventh Marcel Grossman Meeting on
  General Relativity}} (\bibinfo{publisher}{World Scientific},
  \bibinfo{address}{Singapore}, \bibinfo{year}{1995}).

\bibitem[{\citenamefont{Bunkowski et~al.}(2004)\citenamefont{Bunkowski,
  Burmeister, Beyersdorf, Danzmann, Schnabel, Clausnitzer, Kley, and
  T{\"u}nnermann}}]{bunki1}
\bibinfo{author}{\bibfnamefont{A.}~\bibnamefont{Bunkowski}},
  \bibinfo{author}{\bibfnamefont{O.}~\bibnamefont{Burmeister}},
  \bibinfo{author}{\bibfnamefont{P.}~\bibnamefont{Beyersdorf}},
  \bibinfo{author}{\bibfnamefont{K.}~\bibnamefont{Danzmann}},
  \bibinfo{author}{\bibfnamefont{R.}~\bibnamefont{Schnabel}},
  \bibinfo{author}{\bibfnamefont{T.}~\bibnamefont{Clausnitzer}},
  \bibinfo{author}{\bibfnamefont{E.-B.} \bibnamefont{Kley}}, \bibnamefont{and}
  \bibinfo{author}{\bibfnamefont{A.}~\bibnamefont{T{\"u}nnermann}},
  \bibinfo{journal}{Opt. Lett.} \textbf{\bibinfo{volume}{29}},
  \bibinfo{pages}{2342} (\bibinfo{year}{2004}).

\bibitem[{\citenamefont{Bunkowski et~al.}(2005)\citenamefont{Bunkowski,
  Burmeister, Danzmann, and Schnabel}}]{bunki2}
\bibinfo{author}{\bibfnamefont{A.}~\bibnamefont{Bunkowski}},
  \bibinfo{author}{\bibfnamefont{O.}~\bibnamefont{Burmeister}},
  \bibinfo{author}{\bibfnamefont{K.}~\bibnamefont{Danzmann}}, \bibnamefont{and}
  \bibinfo{author}{\bibfnamefont{R.}~\bibnamefont{Schnabel}},
  \bibinfo{journal}{Opt. Lett.} \textbf{\bibinfo{volume}{30}},
  \bibinfo{pages}{1183} (\bibinfo{year}{2005}).

\bibitem[{\citenamefont{Heinzel}(1999)}]{2mc}
\bibinfo{author}{\bibfnamefont{G.}~\bibnamefont{Heinzel}}, Ph.D. thesis,
  \bibinfo{school}{Universit{\"a}t Hannover} (\bibinfo{year}{1999}),
  \bibinfo{note}{internal report MPQ 243 p.207ff}.

\bibitem[{\citenamefont{Mizuno et~al.}(1997)\citenamefont{Mizuno, R{\"u}diger,
  Schilling, Winkler, and Danzmann}}]{miz1}
\bibinfo{author}{\bibfnamefont{J.}~\bibnamefont{Mizuno}},
  \bibinfo{author}{\bibfnamefont{A.}~\bibnamefont{R{\"u}diger}},
  \bibinfo{author}{\bibfnamefont{R.}~\bibnamefont{Schilling}},
  \bibinfo{author}{\bibfnamefont{W.}~\bibnamefont{Winkler}}, \bibnamefont{and}
  \bibinfo{author}{\bibfnamefont{K.}~\bibnamefont{Danzmann}},
  \bibinfo{journal}{Opt. Comm.} \textbf{\bibinfo{volume}{138}},
  \bibinfo{pages}{383} (\bibinfo{year}{1997}).

\bibitem[{\citenamefont{Winkler}(1991)}]{winkblair}
\bibinfo{author}{\bibfnamefont{W.}~\bibnamefont{Winkler}},
  \emph{\bibinfo{title}{The Detection of gravitational waves}}
  (\bibinfo{publisher}{Cambridge University Press},
  \bibinfo{address}{Cambridge}, \bibinfo{year}{1991}), \bibinfo{note}{p.269}.

\bibitem[{\citenamefont{Mizuno}(1995)}]{miz3}
\bibinfo{author}{\bibfnamefont{J.}~\bibnamefont{Mizuno}}, \bibinfo{type}{Tech.
  Rep.} \bibinfo{number}{MPQ 203}, \bibinfo{institution}{Max-Planck-Institut
  f{\"u}r Quantenoptik}, \bibinfo{address}{Garching} (\bibinfo{year}{1995}).

\end{thebibliography}

\end{document}